\documentclass[twocolumn,preprintnumbers,amsmath,amssymb,aps, prb,floatfix,groupedaddress]{revtex4}
\usepackage{txfonts}
\usepackage{graphicx} % Include figure files
\usepackage{subfigure,diagbox}
\usepackage{dcolumn} % Align table columns on decimal point
\usepackage{bm} % Allow bold math
\usepackage{epsfig}
\usepackage{color}
\usepackage{multirow,longtable,textgreek,upgreek}

\begin{document}

\title{Contrast of LiFeAs with isostructural, isoelectronic,\\ and non-superconducting MgFeGe}

\author{H. B. Rhee}
\author{W. E. Pickett}
\affiliation{Department of Physics, University of California, Davis, CA 95616, USA}

\date{February 11, 2013}

\begin{abstract}
Stoichiometric LiFeAs at ambient pressure is an 18 K superconductor while isostructural, isoelectronic MgFeGe does not superconduct, despite their extremely similar electronic structures. To investigate possible sources of this distinctively different superconducting behavior, we quantify the differences using first principles density functional theory. Total Fe $3d$ occupations are identical, with individual $3d$ orbital occupations differing by no more than 0.015. However, a redistribution of bands just above the Fermi level $\varepsilon_F$ provides an important distinction, with more Fe-derived states within 0.5 eV of the Fermi level and a higher $N(\varepsilon_F$) in MgFeGe. For many mechanisms these features would enhance the tendency toward superconductivity by providing more Cooper pairs (in MgFeGe), but the tendency toward magnetic instability might be more important.  Two of the five Fermi surfaces differ between LiFeAs and MgFeGe, but still lead to similar $q$-dependencies of susceptibilities $\chi_0(\mathbf q)$ including the familiar broad peak at $(\pi,\pi)$. The larger $\chi_0(\mathbf q)$ in MgFeGe, by  10-15\% throughout the zone, leads us to tentatively identify this proximity to magnetic instability as the feature underlying the absence of superconductivity in MgFeGe.  Another significant difference is the 2.5\% difference of the in-plane lattice constant, positioning LiFeAs as a chemically compressed version of MgFeGe. This has possible significance since certain Fe pnictides display pressure-induced superconductivity.

\end{abstract}
\maketitle

\section{Introduction}
In the four years since the discovery of superconductivity in the iron-pnictide and -chalcogenide superconductors (FeSCs), a great deal has been learned about the materials physics of the handful of structural subclasses that comprise these new high temperature superconductors.  Differences between the subclasses have been uncovered, but progress on the understanding of the microscopic mechanism of pairing is lacking.  We suppose (as is commonly held) that the superconductivity that occurs in these materials which have in common a layered, fluorite-type Fe-Pn backbone (Pn = pnictide or chalcogenide) has a common origin, at least for the ``high $T_c$'' ($> 10$--15~K) cases. One of the greatest current needs is to identify microscopic characteristics that can shed light specifically on the existence, or not, of high temperature superconductivity (HTS) and thereby on the underlying pairing mechanism.

Several mechanisms have been put forward.  Because superconductivity
borders and competes with magnetic order as in the HTS cuprates, it is natural to study a spin fluctuation (SF) origin, and several groups\cite{fuseya2009,zhai2009,maier2011,fang2011} have pursued antiferromagnetic SF models. Fe~$3d$ orbital occupation and character have received much attention, and an orbital fluctuation (OF) model has been suggested by Saito \textit{et al.}\cite{saito2010}  The role of the Pn anion was given more attention in the charge fluctuation (CF) picture of Zhou \textit{et al.},\cite{zhou2011} where interatomic Fe-Pn charge-charge interactions provided another electronic mechanism.  On one hand, strong mixing with the narrow Fe~$3d$ bands removes almost all Pn character in the states near the Fermi level
$\varepsilon_F$, making most models focus simply on the $3d$ states. On the
other hand, $T_c$ has been found to correlate strongly with the distance of the Pn above and below the Fe plane, or more specifically on the Pn-Fe-Pn bond angle and/or Fe-Pn bond length.\cite{lee,johrendt}

The Pn anion received attention early in the study of FeSCs, when Berciu \textit{et al.}\cite{berciu2009} modeled pairing in terms of electronic polarons and bipolarons. Their picture foreshadowed the CF model mentioned above, but placed more emphasis on the anticipated large polarizability of the Pn anions, and did not include consideration of SFs or OFs. This picture was continued and extended to include interatomic exchange coupling by Chan \textit{et al.}\cite{chan2010} The conventional phonon mechanism, for which there is a reliable microscopic theory when electronic interactions are described sufficiently by density functional methods, has been evaluated by Boeri \textit{et al.}\cite{boeri2008} and found to be too weak to explain $T_c$ in the 25--55~K range. Strong spin-lattice coupling, however, has been suggested by Egami \textit{et al.}\cite{egami2010} to be involved in pairing.
The electronic structure of these FeSCs is being studied in detail experimentally and modeled carefully by numerous groups, with a primary aim being to uncover the pairing mechanism. 

LiFeAs is recognized as a problem child in the categorization of FeSCs.
Among the vast collection of FeSCs that have been discovered since the first of their kind,\cite{kamihara} LiFeAs is one of very few which superconduct without the need for either chemical doping or physical compression. Of these few, LiFeAs not only has the highest superconducting transition temperature ($T_c=18$ K), but it is to date the only compound, other than LiFeP, whose $T_c$ does {\it not} increase when doped or pressurized. It is widely believed that in the majority of FeSCs, doping or application of pressure suppresses the nesting-induced spin-density-wave (SDW) order to make way for competing superconductivity. LiFeAs however does not undergo any magnetic transition, and significance of FS nesting in LiFeAs has been questioned.\cite{borisenko} It also differs from its isovalent sister compound NaFeAs. Similarities between NaFeAs and LiFeAs in band structure and DOS are even more pronounced\cite{kusakabe,deng} than between MgFeGe and LiFeAs, but unlike MgFeGe and LiFeAs, NaFeAs undergoes a magnetically driven structural phase transition above the superconducting transition.\cite{he,ma,kitagawa} In fact, bulk superconductivity may not even exist in NaFeAs.\cite{parker,chu,morozov}

The recent report of the synthesis and characterization by Hosono's group\cite{mgfege} of non-magnetic MgFeGe, which is isostructural and isoelectronic with the 18~K superconductor LiFeAs but is not superconducting, provides a means to obtain new insight. In their initial report, Liu \textit{et al.}\cite{mgfege} noted the resemblance of the electronic structure near $\varepsilon_F$ to that of LiFeAs, providing both a conundrum and an opportunity to identify differences that account for the vast distinction in superconducting behavior. Supposing that Mg gives up both of its valence electrons to the Fe-Ge bands, two initial basic questions emerge: the $3d$ charge on Fe and its orbital distribution, and the distinctions between the anions As and Ge that are neighbors in the periodic table. 

We first establish the unexpected feature that the Fe~$3d$ occupation is identical in these two compounds. 
The (presumably) more negatively charged Ge anion should have an even higher polarizability than As, for which the simplest viewpoint might suggest to be more \textit{favorable} for superconductivity, rather than precluding it. Furthermore, the As-Fe-As angle, which has been shown for most of the pnictides\cite{lee,johrendt} to correlate strongly with increased $T_c$ as it approaches the regular tetrahedral angle of 109.47$^\circ$, is 103.1$^\circ$, very similar to MgFeGe's Ge-Fe-Ge angle of 103.6$^\circ$. Note that NaFeAs, which possesses an angle of 108.3$^\circ$, is a departure from the geometry of these two compounds in this respect, as well as from the general trends among the FeSCs. One can also question whether the alkali versus alkali earth atom can make any real difference for superconductivity.  Otherwise, the essential difference must come down to small distinctions on the Fe atom, such as individual orbital occupations or details of the band structure which, as we show, are quite similar in these compounds.

In this paper we perform a close comparison of the electronic structures of MgFeGe and LiFeAs. The differences of the Fe~$3d$ orbital occupations, though small as mentioned above, are however readily quantifiable. The DOS near and at the Fermi energy is a major difference between LiFeAs and MgFeGe.  Electronic susceptibility calculations show that both compounds behave similarly overall, with structure closely related to that calculated and observed in 1111 and 122 FeSCs. A summary of findings is provided in the last section.

%---------------------------------------------------------------------
\section{Electronic structure}

\subsection{Description of methods}

\begin{table}[tb]
\caption{Crystal structure parameters of LiFeAs and MgFeGe in the tetragonal space group $P4/nmm$. Wyckoff labels of Li/Mg, Fe, and As/Ge are, respectively, 2c, 2b, and 2c.}
%\begin{ruledtabular}
\begin{tabular}{lcc|ll} 
& $a$ (\AA)   & $c$ (\AA) &    & $(x,y,z)$\\
\hline\hline
\multirow{3}{*}{LiFeAs}
&             &           & Li & $(1/4, 1/4, 0.8459)$ \\
& 3.7914      & 6.3639    & Fe & $(3/4, 1/4,    1/2)$ \\
&             &           & As & $(1/4, 1/4, 0.2635)$ \\
\hline
\multirow{3}{*}{MgFeGe}
&             &           & Mg & $(1/4, 1/4, 0.8316)$ \\
& 3.8848      & 6.4247    & Fe & $(3/4, 1/4,    1/2)$ \\
&             &           & Ge & $(1/4, 1/4, 0.2620)$ \\
\hline
\end{tabular}
\label{tbl:xtal}
%\end{ruledtabular}
\end{table}

The non-magnetic electronic structures of LiFeAs and MgFeGe were calculated using {\sc WIEN2k},\cite{wien,wien2} a density functional theory (DFT)-based full potential, linearized augmented planewave (LAPW) code. For the exchange-correlation functional, the generalized gradient approximation (GGA) of Perdew, Burke, and Ernzerhof\cite{pbe} was applied. A $k$-point mesh of $32 \times 32 \times 19$ was used, and $RK_{max}$ was set to 10. As for the muffin-tin (MT) radii $R$, we used values of 2.50, 2.43, and 2.14~a.u.\ for Li/Mg, Fe, and As/Ge, respectively. We adopted the experimental lattice constants of LiFeAs and MgFeGe (both of which form a \textit{P}4/\textit{nmm} tetragonal unit cell) given in Refs.~\citenum{mgfege} and \citenum{tapp} respectively. The internal coordinates came from the same experiments, and relaxation was not applied in our calculations since we are interested in identifying distinctions and comparing with experiment when possible. The structural parameters are given in Table~\ref{tbl:xtal}.  We have verified that changes due to spin-orbit coupling are uninterestingly tiny. Note that the Li (Mg) atom is five-coordinate with As (respectively, Ge) in a square-based pyramid, and its height above the Fe plane is 2.20~\AA\ (2.13~\AA).

\begin{figure}[bt]
\begin{center}
\includegraphics[draft=false,width=\columnwidth,clip]{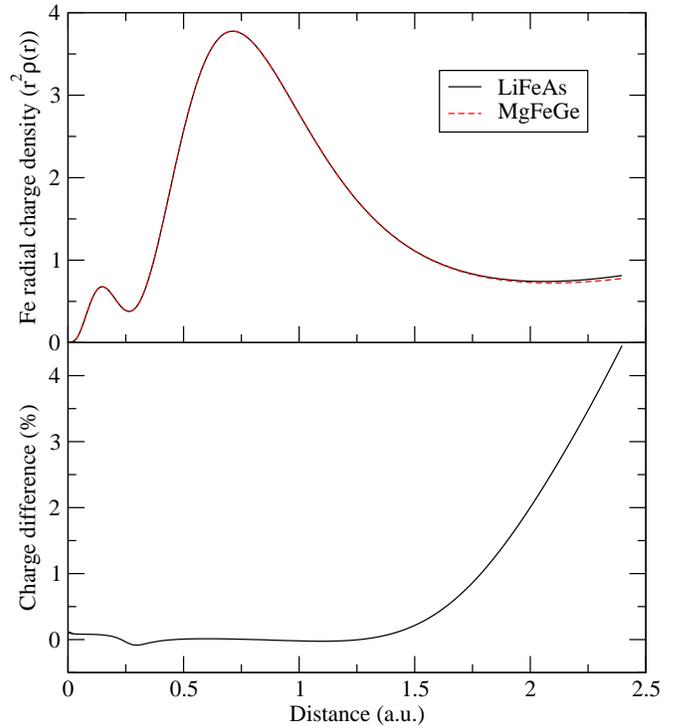}
\end{center}
\caption{(Color online) Valence radial charge density (in a.u.) of Fe in LiFeAs and MgFeGe (top), illustrating the extreme similarity not only in the region of the $3d$ peak at 0.7 a.u., but also extending out to 1.4 a.u. The tiny difference is quantified by plotting the percentage difference (bottom). Both are plotted versus the radial distance from the Fe atom center.}
\label{rad}
\end{figure}

\subsection{Charge density and As/Ge-related differences}
The Fe~$3d$ occupations in LiFeAs and MgFeGe are identical. This essential feature is established by comparing the radial valence charge density $4\pi r^2\rho(r)$ of Fe between the two compounds, shown in Fig.~\ref{rad}. In the region of the $3d$ peak at radius $r=0.7$~a.u., where the density is due only to $3d$ occupation and to identical core tails, the difference is incredibly small (less than 0.1\%), and the difference remains tiny out to 1.4~a.u. Using the same MT radius of 2.43 a.u., the charge of Fe in this sphere in LiFeAs is larger only by 0.01 electrons (the charges are 6.65 for LiFeAs, 6.64 for MgFeGe), with this tiny difference arising from the differing As and Ge tail charges. The orbital occupation matrix elements $n_{mm'}$ for both compounds are listed for comparison in Table~\ref{tbl:occs}. The largest difference is 0.015 for the $d_{z^2}$ ($m=0$) orbital, being smaller in MgFeGe. This difference is compensated by the $d_{xz/yz}$ occupation ($m=\pm1$) being larger by $\sim 0.01$ in MgFeGe. The $m=\pm2$ ($d_{xy}$, $d_{x^2-y^2}$) orbital occupation differences are negligible. The $t_{2g}$ and $e_g$ degeneracies are broken by the Fe site symmetry, of course, as can be noted in the occupation matrices.

\begin{table}[bt]
\centering
\subtable{
\begin{tabular}{|r|rrrrr|}
\multicolumn{6}{c}{LiFeAs} \\
\hline
\diagbox{$m$}{$m'$} & -2 & -1 & 0 & +1 & +2\\ \hline
-2 & 0.646 & 0.000 & 0.000 & 0.000 &-0.033  \\
-1 & 0.000 & 0.626 & 0.000 & 0.000 & 0.000  \\
 0 & 0.000 & 0.000 & 0.698 & 0.000 & 0.000  \\
+1 & 0.000 & 0.000 & 0.000 & 0.616 & 0.000  \\
+2 &-0.033 & 0.000 & 0.000 & 0.000 & 0.625  \\ \hline
\end{tabular}
}
\vspace{10pt}
\subtable{
\begin{tabular}{|r|rrrrr|}
\multicolumn{6}{c}{MgFeGe} \\
\hline
\diagbox{$m$}{$m'$} & -2 & -1 & 0 & +1 & +2\\ \hline
-2 & 0.645 & 0.000 & 0.000 & 0.000 &-0.038  \\
-1 & 0.000 & 0.635 & 0.000 & 0.000 & 0.000  \\
 0 & 0.000 & 0.000 & 0.683 & 0.000 & 0.000  \\
+1 & 0.000 & 0.000 & 0.000 & 0.623 & 0.000  \\
+2 &-0.038 & 0.000 & 0.000 & 0.000 & 0.621  \\ \hline
\end{tabular}
}
\caption{Fe~$3d$ orbital occupation matrix elements $n_{mm'}$ of LiFeAs and MgFeGe, from the LAPW sphere of radius $R=2.47$~a.u.}
\label{tbl:occs}
\end{table}

Since the total Fe~$3d$ occupation is identical and individual orbital occupations differ by only 1--2\% for the two compounds, it becomes of more interest to compare differences ascribable to As and Ge. One difference is the structure itself: the lattice constants $a$ and $c$ of MgFeGe are about 2.5\% and 1\% larger respectively, giving MgFeGe a 6\% larger volume. Of Fe's three nearest-neighbor interatomic distances, the Fe-Fe distance differs most, 2.68~\AA\ compared to 2.75~\AA, this 2.5\% increase being directly related to the same relative increase in $a$ lattice constant. In this sense LiFeAs is a compressed version of MgFeGe. Effective $d$-$d$ hopping amplitudes might be changed somewhat, however, since the hopping is largely through the As or Ge atom and this interaction will differ. There is a similar 2\% increase in the Fe-As/Ge distance (2.42~\AA/2.47~\AA), related to Ge's larger atomic radius. Despite MgFeGe's larger unit cell (6\% in volume), and Mg having larger nuclear charge than Li, but perhaps because Mg has an additional valence electron, the Fe-Mg distance is 0.8\% \textit{shorter} than that of Fe-Li (2.91~\AA\ versus 2.88~\AA).

There has been interest in the polarizability of the pnictide atom and its effect on carriers. The polarizability of the Ge atom is 40\% larger than that of As (6.07 vs.\ 4.31 \AA$^3$). Naively, this difference would favor superconductivity in MgFeGe according to the relevant model.\cite{sawatzky} Moreover, with the additional valence electron (Mg vs.\ Li) not going to Fe and thus going primarily to Ge, this difference in polarizabilities of Ge and As may be enhanced in the solid, making it even more difficult to understand the difference in superconducting behavior in terms of the metalloid polarizability.

\begin{figure}[tb]
  \mbox{\begin{minipage}[t]{\columnwidth}
    \includegraphics[width=\columnwidth]{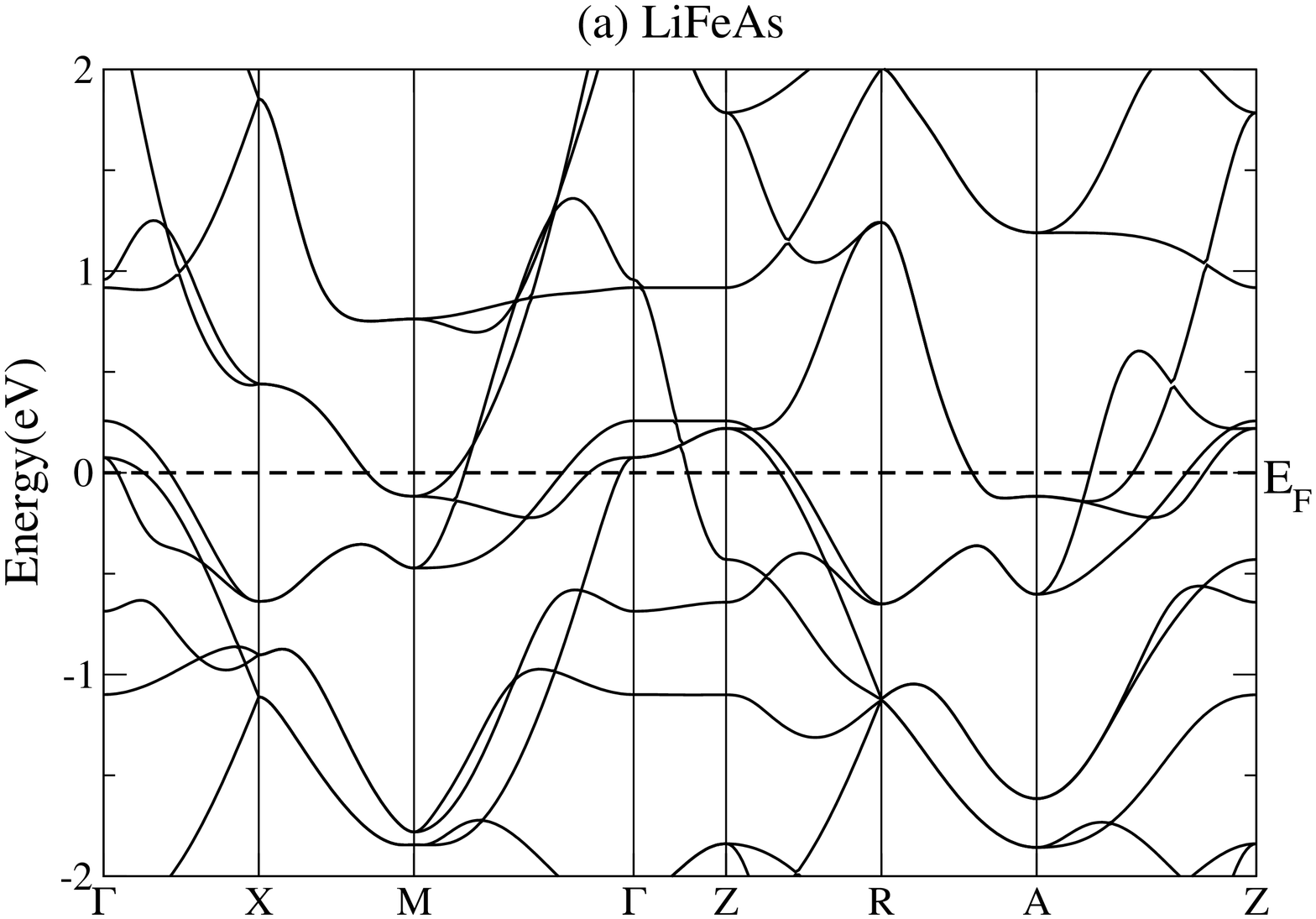}
    \vspace{5pt}
  \end{minipage}}
  \hfill
  \mbox{\begin{minipage}[t]{\columnwidth}
    \includegraphics[width=\columnwidth,clip]{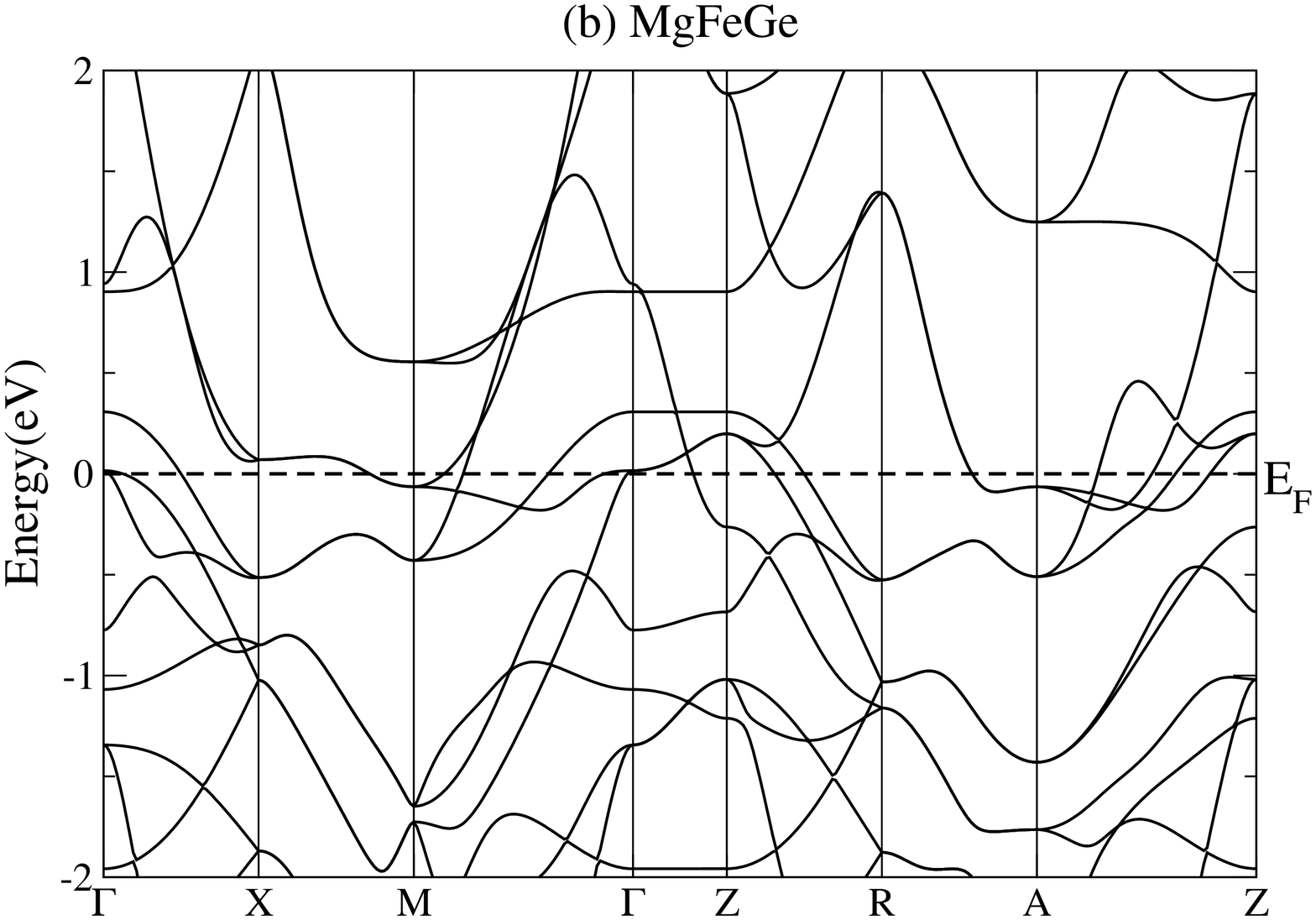}
  \end{minipage}}
\caption{Band structures of LiFeAs and MgFeGe near the Fermi energy. The differences at $\Gamma$, X, and M are discussed in the text.}
\label{bs}
\end{figure}

\subsection{Bands and Fermi surfaces}

The band structures of the compounds are compared in Fig.~\ref{bs}. As noted by Liu \textit{et al.}, the similarity of their bands near the Fermi energy $\varepsilon_F$ is a striking feature. The doubly degenerate band at $\Gamma$ drops by less than 0.1~eV to lie very close to the Fermi level in MgFeGe, and the pocket at M shrinks in size due to a rise by less than 0.1~eV of the band just below $\varepsilon_F$; the bands elsewhere at $\varepsilon_F$ are extremely similar along symmetry lines. However, the lowering by 0.38~eV (from 0.43~eV to 0.05~eV) of the lowest unoccupied band at the X point is a comparatively drastic difference.

\begin{figure}[bt]
\begin{center}
\includegraphics[draft=false,width=\columnwidth]{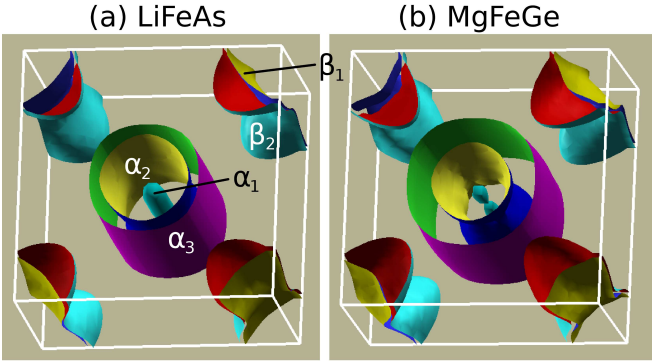}
\end{center}
\caption{(Color online) Fermi surfaces, in the Brillouin zone, of (a) LiFeAs and (b) MgFeGe. In both cases, three hole pockets (\textalpha$_1$, \textalpha$_2$, \textalpha$_3$) center $\Gamma$, located at the unit center, and two electron pockets (\textbeta$_1$, \textbeta$_2$) surround each M and A point, the latter located at the zone corners.}
\label{fss}
\end{figure}

The FSs of FeSCs are believed to be intimately connected to their superconducting properties and magnetic excitation spectrum. The FSs of LiFeAs and MgFeGe, depicted in Fig.~\ref{fss}, are each made up of three hole pockets centered at $\Gamma$, denoted \textalpha$_1$, \textalpha$_2$, and \textalpha$_3$ from smaller to larger, and two electron pockets at M, \textbeta$_1$ and \textbeta$_2$. The largest hole pocket \textalpha$_3$ of MgFeGe is larger than that of LiFeAs, while the two smaller hole-like pockets \textalpha$_1$ and \textalpha$_2$ are smaller. The waists (at $k_z=0$) of both \textbeta$_1$ and \textbeta$_2$ are smaller---the corresponding band lies almost exactly at $\varepsilon_F$---but MgFeGe's pockets encompass more electrons than does LiFeAs due to their more extreme flaring around the zone corner A. \textalpha$_1$ and \textalpha$_2$ are primarily $d_{xz/yz}$-like, and \textalpha$_3$ is mainly $d_{xy}$ in character. At $k_z=0$, \textbeta$_1$ and \textbeta$_2$ have, respectively, $d_{xz/yz}$ and $d_{xy}$ character, but as $k_z \rightarrow \pm\pi$, they switch attributes so that \textbeta$_1$ becomes predominantly $d_{xy}$-like and \textbeta$_2$ mostly $d_{xz/yz}$-like at $k_z=\pi$.

\subsection{Density of states}

\begin{figure}[bt]
\begin{center}
\includegraphics[draft=false,width=\columnwidth,clip]{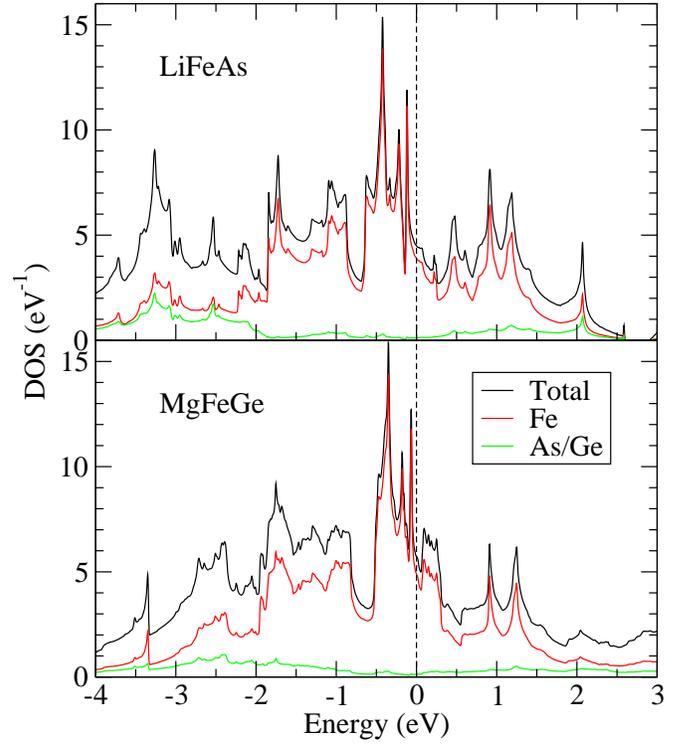}
\end{center}
\caption{(Color online) Densities of states (per two spins, per primitive cell) of LiFeAs and MgFeGe. Li/Mg states give negligible contribution in this energy range so they have not been plotted. The primary difference occurs in the 0.0--0.5~eV regions (see text).}
\label{dos}
\end{figure}

The densities of states (DOSs) of LiFeAs and MgFeGe are aligned for comparison in Fig.~\ref{dos}. In the case of LiFeAs, $\varepsilon_F$ sits on the shoulder of a high DOS region with almost entirely Fe~$d$ character, consistent with results presented by previous groups.\cite{singh,nekrasov,mgfege} Li/Mg contributions are negligible at and around the Fermi energy, and As/Ge density is also very low at $\varepsilon_F$. The LiFeAs and MgFeGe DOSs share similar Fe~$d$ features from $-2$~eV up to $\varepsilon_F$, although the latter has a somewhat compressed range of large DOS. Above $\varepsilon_F$ in MgFeGe the shoulder is more abruptly cut off. But the peak of unoccupied states centered at 0.5~eV in LiFeAs is shifted down in MgFeGe to form a denser manifold just above the abridged shoulder, and it is centered 0.2~eV above the Fermi level. This can be traced in part to the drop in the first unoccupied band at X, mentioned earlier, providing more states near the Fermi energy.
The DOS at $\varepsilon_F$, $N(0)$, is 4.5~eV$^{-1}$ in LiFeAs (per unit cell of two formula units), which is comparable to that of many other FeSCs.\cite{shein,singh2,leithe,shein2,neupane,nekrasov2} In MgFeGe $N(0)$ = 5.7~eV$^{-1}$, 20\% larger than in LiFeAs. 

A higher $N(0)$, and more generally more states within 0.5 eV of $\varepsilon_F$, is a property that conventionally should enhance superconductivity, since $N(0)$ is the measure of the number of condensed Cooper pairs. However in several FeSCs, the superconducting state has a lower DOS than in its parent (normal) phase.\cite{goldman,mazin,xu,leithe} As mentioned earlier, most undoped parent compounds of FeSCs do not superconduct at ambient pressure; they have to be compressed, or electrons or holes have to be introduced, in order to influence them into the superconducting regime. Pressure should, generally speaking, decrease the DOS at the Fermi energy, since bands widen with spatial compression. This so happens in CaFe$_2$As$_2$ as it transitions from the non-superconducting state to the compressed superconducting phase.\cite{goldman} As for chemical doping, two groups\cite{mazin,xu} report a drop in $N(0)$ when non-superconducting LaOFeAs, with F doping, becomes superconducting; SrFe$_2$As$_2$ also has a smaller DOS value at the Fermi energy when Co is introduced into the system to trigger superconductivity;\cite{leithe} and BaFe$_2$As$_2$, when electron-doped, passes over to the superconducting phase while suppressing magnetism,\cite{canfield,sefat,ning} and $N(0)$ is indeed expected to decrease. In each case of all of these compounds, the weak magnetic ordering exhibited by the parent compound is gradually lost as superconductivity is turned on. Typically the role of doping or pressurization in the FeSCs is to remove magnetic instability, which can be done by lowering the DOS at the Fermi energy.

Angle-resolved photoemission spectroscopy (ARPES) measurements\cite{borisenko,lankau,umezawa,hajiri,skornyakov,gslee} have shown differences when compared to the DFT band structure of LiFeAs, with some DFT bands crossing $\varepsilon_F$ being too high in energy and others being too low. Along the symmetry lines $\Gamma$-X-M-$\Gamma$, at every $k$ point where a band in ARPES\cite{borisenko} crosses the Fermi energy, the DFT inaccuracy is 70~meV or less. But at every energy within $\varepsilon_F$$\pm 70$~meV, the DOS of LiFeAs is lower than that of MgFeGe. Moreover, MgFeGe has 40\% more electronic states in the $\varepsilon_F$$\pm$ 70 meV window than LiFeAs. Thus, despite the impreciseness of the DFT band structures, it is clear that MgFeGe will have a higher $N(0)$. We suggest that the larger $N(0)$ in MgFeGe, versus LiFeAs, can explain the absence of superconductivity: MgFeGe is closer to the SDW instability, further from the superconducting region of the phase diagram. 

\subsection{Susceptibility $\chi_0(\mathbf q)$}

The ARPES studies referred to above are universal in their agreement that FSs \textalpha$_1$ and \textalpha$_2$ are considerably smaller (by about 95\% and 80\%, respectively) than their DFT analogs. Many published papers have thus drawn the conclusion that nesting would be completely absent or much weaker than that in the 1111 and 122 Fe-based superconductors, whose hole and electron pockets are more equal in size. It has, however, been pointed out that structure in $\chi_0 (\mathbf q)$ is not drawn solely from the FS, but can be affected by bands somewhat away from the FS.\cite{johannes} Thus ``nesting'' can be ambiguous until its use is clarified.

\begin{figure}[tb]
\begin{center}
\includegraphics[draft=false,width=0.8\columnwidth]{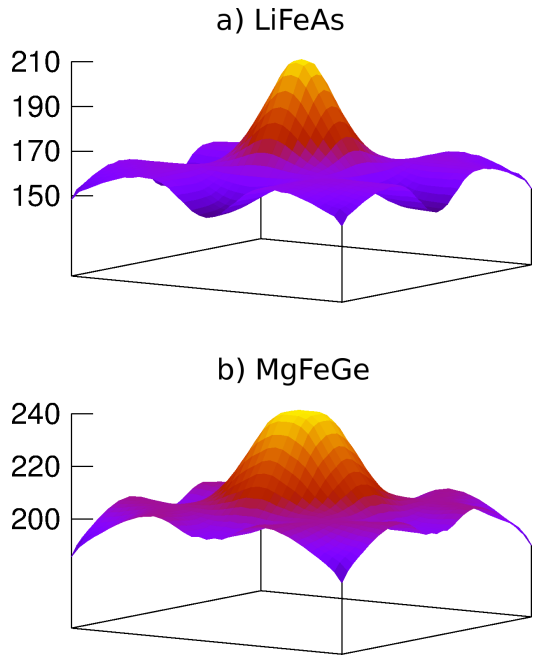}
\end{center}
\caption{(Color online) Non-interacting spin susceptibilities $\chi_0(q_x,q_y,q_z\equiv 0)$, in arbitrary units, of (a) LiFeAs and (b) MgFeGe in the complete Brillouin zone. A distinctive difference is that the peak at $(\pi,pi,0)$ is higher but wider in MgFeGe. In these plots, the $\Gamma$ point is positioned at the corners.}
\label{susc}
\end{figure}

In Figure~\ref{susc}(a) and (b), we provide the calculated constant-matrix-element bare susceptibility $\chi_0(\mathbf q)$ (in the same arbitrary but comparable units) of LiFeAs and MgFeGe, respectively, obtained from all five bands that cross the Fermi level. Our $\chi_0$ for LiFeAs is similar in shape to that calculated by Lee \textit{et al.}.\cite{gslee} There is a strong but broad maximum in $\chi_0$ at $(\pi,\pi)$ in both materials, and the two $\Gamma$-centered cylindrical FSs with radii $k_{F,\upalpha_2}$ and $k_{F,\upalpha_3}$ lead to the circular hump in the range of $2k_{F,\upalpha_2}$, $2k_{F,\upalpha_3}$, and $k_{F,\upalpha_2}$ + $k_{F,\upalpha_3}$. The susceptibility of LiFeAs in fact resembles that of superconducting LaO$_{0.9}$F$_{0.1}$FeAs calculated by Mazin \textit{et al.}\cite{mazin} Though Dong \textit{et al.}\cite{dong} find the peak at M to collapse with F doping, they do agree with Mazin \textit{et al.} that the susceptibility is stronger in the parent compound. It is not however the smaller hole pockets but the largest pocket that predominantly contributes to the peak in $\chi_0(\mathbf q)$. We have confirmed this by calculating $\chi_0$ excluding the two hole pockets, finding that its topology is virtually identical to that which includes all pockets.

The larger magnitude of $\chi_0(\pi,\pi)$ of MgFeGe compared to LiFeAs, which is related to MgFeGe's larger $N(0)$ value, supports the evidence provided above that MgFeGe is closer to a magnetic instability, plausibly accounting for its lack of superconductivity. Three studies\cite{yin,ferber,gslee} have shown that applying dynamical mean-field theory (DMFT), in which dynamical electron correlation effects are taken into account in a local manner, yields FSs more consistent with experiment for LiFeAs. According to Lee \textit{et al.}'s calculations, DMFT reduces the $(\pi,\pi)$ peak but not completely ---$\chi_0(\mathbf q(\pi,\pi))/\chi_0(\mathbf q=0)$ drops below 1.1---and broadening remains the same. Application of DMFT to MgFeGe, as well as ARPES measurements on the material, will be instructive in trying to understand differences underlying the different superconducting behavior of these two very similar compounds.

\section{Summary}

In this study we have made a comparison of several aspects of the electronic structures of LiFeAs and isoelectronic but nonsuperconducting MgFeGe. As noted in the original report,\cite{mgfege} the band structures near the Fermi level are very similar. Moreover the Ge-Fe-Ge and As-Fe-As bond angles, which correlate strongly with T$_c$ across the classes of FeSCs, are nearly the same and thus violate the general trend. We have determined that the Fe $3d$ occupation is also identical for the two compounds.

Several differences have been identified. Individual Fe $3d$ orbital occupations differ by up to $0.015$,  mounting to relative changes of 0--2\%. A repositioning of a DOS peak a few tenths of eV above $\varepsilon_F$, and an overall increase in the Fe $3d$ DOS within 0.5 eV of $\varepsilon_F$, are the most obvious differences, reflecting differences in Fe hybridization with Ge versus As. Judging from the sizes of Fermi surfaces the degree of nesting changes, but the same broad but significant peak of the susceptibility seen in other magnetically suppressed FeSCs exists in LiFeAs as well. The shape of $\chi_0(\mathbf q)$ is similar in MgFeGe but the intensity is greater, a feature that figures ito spin fluctuation scenarios of FeSCs. 

The higher DOS of MgFeGe at and near $\varepsilon_F$, as well as the larger $\chi_0$ throughout the zone and 15\% larger at $(\pi,\pi)$, implies that it is more proximate to a magnetic instability, which we tentatively identify as the most likely factor in the absence of superconductivity. Another observation is that the smaller in-plane lattice constant in LiFeAs (5\% smaller unit cell area) positions it as a strongly compressed version of MgFeAs, also impacting their different superconducting behaviors. We suggest that further exploration in parallel of these two compounds, theoretically and by experiment, will provide one of the most promising approaches to identifying the superconducting mechanism in FeSCs.

%\acknowledgment
\section*{Acknowledgment}
We thank Y.~Quan for emphasizing the importance of the radial density in characterizing the $3d$ occupation in transition metal compounds.  
This work was supported by a DOE Computational Materials and Chemical Science Network grant DE-SC0005468.

\bibliography{refs}

\begin{thebibliography}{10}

\bibitem{fuseya2009}
Y.~Fuseya, T.~Kariyado, and M.~Ogata: J.~Phys. Soc. Jpn. {\bfseries 78} (2009)
  023703.

\bibitem{zhai2009}
H.~Zhai, F.~Wang, and D.-H. Lee: Phys. Rev. B {\bfseries 80} (2009) 064517.

\bibitem{maier2011}
T.~A. Maier, S.~Graser, P.~J. Hirschfeld, and D.~J. Scalapino: Phys. Rev. B
  {\bfseries 83} (2011) 100515.

\bibitem{fang2011}
C.~Fang, Y.-L. Wu, R.~Thomale, B.~A. Bernevig, and J.~Hu: Phys. Rev. X
  {\bfseries 1} (2011) 011009.

\bibitem{saito2010}
T.~Saito, S.~Onari, and H.~Kontani: Phys. Rev. B {\bfseries 82} (2010) 144510.

\bibitem{zhou2011}
S.~Zhou, G.~Kotliar, and Z.~Wang: Phys. Rev. B {\bfseries 84} (2011) 140505.

\bibitem{lee}
C.-H. Lee, A.~Iyo, H.~Eisaki, H.~Kito, M.~T. Fernandez-Diaz, T.~Ito, K.~Kihou,
  H.~Matsuhata, M.~Braden, and K.~Yamada: J.~Phys. Soc. Jpn. {\bfseries 77}
  (2008) 083704.

\bibitem{johrendt}
D.~Johrendt, H.~Hosono, R.-D. Hoffmann, and R.~P\"ottgen: Z. Kristallogr.
  {\bfseries 226} (2011) 435.

\bibitem{berciu2009}
M.~Berciu, I.~Elfimov, and G.~A. Sawatzky: Phys. Rev. B {\bfseries 79} (2009)
  214507.

\bibitem{chan2010}
H.~Takahashi, Y.~Imai, S.~Komiya, I.~Tsukada, and A.~Maeda: Phys. Rev. B
  {\bfseries 84} (2011) 132503.

\bibitem{boeri2008}
L.~Boeri, O.~V. Dolgov, and A.~A. Golubov: Phys. Rev. Lett. {\bfseries 101}
  (2008) 026403.

\bibitem{egami2010}
T.~Egami, B.~Fine, D.~Parshall, A.~Subedi, and D.~Singh: Adv. Condens. Matt.
  Phys. {\bfseries 2010} (2010) 164916.

\bibitem{kamihara}
Y.~Kamihara, T.~Watanabi, M.~Hirano, and H.~Hosono: J.~Am. Chem. Soc.
  {\bfseries 130} (2008) 3296.

\bibitem{borisenko}
S.~V. Borisenko, V.~B. Zabolotnyy, D.~V. Evtushinsky, T.~K. Kim, I.~V. Morozov,
  A.~N. Yaresko, A.~A. Kordyuk, G.~Behr, A.~Vasiliev, R.~Follath, and
  B.~B\"uchner: Phys. Rev. Lett. {\bfseries 105} (2010) 067002.

\bibitem{kusakabe}
K.~Kusakabe and A.~Nakanishi: J.~Phys. Soc. Jpn. {\bfseries 78} (2009) 124712.

\bibitem{deng}
S.~Deng, J.~K\"ohler, and A.~Simon: Phys. Rev. B {\bfseries 80} (2009) 214508.

\bibitem{he}
C.~He, Y.~Zhang, B.~P. Xie, X.~F. Wang, L.~X. Yang, B.~Zhou, F.~Chen, M.~Arita,
  K.~Shimada, H.~Namatame, M.~Taniguchi, X.~H. Chen, J.~P. Hu, and D.~L. Feng:
  Phys. Rev. Lett. {\bfseries 105} (2010) 117002.

\bibitem{ma}
L.~Ma, G.~F. Chen, D.-X. Yao, J.~Zhang, S.~Zhang, T.-L. Xia, and W.~Yu: Phys.
  Rev. B {\bfseries 83} (2011) 132501.

\bibitem{kitagawa}
K.~Kitagawa, Y.~Mezaki, K.~Matsubayashi, Y.~Uwatoko, and M.~Takigawa: J.~Phys.
  Soc. Jpn. {\bfseries 80} (2011) 033705.

\bibitem{parker}
D.~R. Parker, M.~J. Pitcher, P.~J. Baker, I.~Franke, T.~Lancaster, S.~J.
  Blundell, and S.~J. Clarke: Chem. Commun.  (2009) 2189.

\bibitem{chu}
C.~Chu, F.~Chen, M.~Gooch, A.~Guloy, B.~Lorenz, B.~Lv, K.~Sasmal, Z.~Tang,
  J.~Tapp, and Y.~Xue: Physica C {\bfseries 469} (2009) 326 .

\bibitem{morozov}
I.~Morozov, A.~Boltalin, O.~Volkova, A.~Vasiliev, O.~Kataeva, U.~Stockert,
  M.~Abdel-Hafiez, D.~Bombor, A.~Bachmann, L.~Harnagea, M.~Fuchs, H.-J. Grafe,
  G.~Behr, R.~Klingeler, S.~Borisenko, C.~Hess, S.~Wurmehl, and B.~B\"uchner:
  Cryst. Growth Des. {\bfseries 10} (2010) 4428.

\bibitem{mgfege}
X.~Liu, S.~Matsuishi, S.~Fujitsu, and H.~Hosono: Phys. Rev. B {\bfseries 85}
  (2012) 104403.

\bibitem{wien}
P.~Blaha, K.~Schwarz, P.~Sorantin, and S.~Trickey: Comp. Phys. Comm. {\bfseries
  59} (1990) 399 .

\bibitem{wien2}
K.~Schwarz and P.~Blaha: Comput. Mater. Sci. {\bfseries 28} (2003) 259 .

\bibitem{pbe}
J.~P. Perdew, K.~Burke, and M.~Ernzerhof: Phys. Rev. Lett. {\bfseries 77}
  (1996) 3865.

\bibitem{tapp}
J.~H. Tapp, Z.~Tang, B.~Lv, K.~Sasmal, B.~Lorenz, P.~C.~W. Chu, and A.~M.
  Guloy: Phys. Rev. B {\bfseries 78} (2008) 060505.

\bibitem{sawatzky}
G.~A. Sawatzky, I.~S. Elfimov, J.~van~den Brink, and J.~Zaanen: Europhys.~Lett.
  {\bfseries 86} (2009) 17006.

\bibitem{singh}
D.~J. Singh: Phys. Rev. B {\bfseries 78} (2008) 094511.

\bibitem{nekrasov}
I.~Nekrasov, Z.~Pchelkina, and M.~Sadovskii: J.~Exp. Theor. Phys. {\bfseries
  88} (2008) 543.

\bibitem{shein}
I.~Shein and A.~Ivanovskii: J.~Exp. Theor. Phys. {\bfseries 88} (2008) 107.

\bibitem{singh2}
D.~J. Singh and M.-H. Du: Phys. Rev. Lett. {\bfseries 100} (2008) 237003.

\bibitem{leithe}
A.~Leithe-Jasper, W.~Schnelle, C.~Geibel, and H.~Rosner: Phys. Rev. Lett.
  {\bfseries 101} (2008) 207004.

\bibitem{shein2}
I.~Shein and A.~Ivanovskii: J.~Exp. Theor. Phys. {\bfseries 88} (2008) 683.

\bibitem{neupane}
M.~Neupane, C.~Liu, S.-Y. Xu, Y.-J. Wang, N.~Ni, J.~M. Allred, L.~A. Wray,
  N.~Alidoust, H.~Lin, R.~S. Markiewicz, A.~Bansil, R.~J. Cava, and M.~Z.
  Hasan: Phys. Rev. B {\bfseries 85} (2012) 094510.

\bibitem{nekrasov2}
I.~Nekrasov and M.~Sadovskii: J.~Exp. Theor. Phys. {\bfseries 93} (2011) 166.

\bibitem{goldman}
A.~I. Goldman, A.~Kreyssig, K.~Proke\ifmmode~\check{s}\else \v{s}\fi{}, D.~K.
  Pratt, D.~N. Argyriou, J.~W. Lynn, S.~Nandi, S.~A.~J. Kimber, Y.~Chen, Y.~B.
  Lee, G.~Samolyuk, J.~B. Le\~ao, S.~J. Poulton, S.~L. Bud'ko, N.~Ni, P.~C.
  Canfield, B.~N. Harmon, and R.~J. McQueeney: Phys. Rev. B {\bfseries 79}
  (2009) 024513.

\bibitem{mazin}
I.~I. Mazin, D.~J. Singh, M.~D. Johannes, and M.~H. Du: Phys. Rev. Lett.
  {\bfseries 101} (2008) 057003.

\bibitem{xu}
G.~Xu, H.~Zhang, X.~Dai, and Z.~Fang: Europhys.~Lett. {\bfseries 84} (2008)
  67015.

\bibitem{canfield}
P.~C. Canfield, S.~L. Bud'ko, N.~Ni, J.~Q. Yan, and A.~Kracher: Phys. Rev. B
  {\bfseries 80} (2009) 060501.

\bibitem{sefat}
A.~S. Sefat, R.~Jin, M.~A. McGuire, B.~C. Sales, D.~J. Singh, and D.~Mandrus:
  Phys. Rev. Lett. {\bfseries 101} (2008) 117004.

\bibitem{ning}
F.~Ning, K.~Ahilan, T.~Imai, A.~S. Sefat, R.~Jin, M.~A. McGuire, B.~C. Sales,
  and D.~Mandrus: J.~Phys. Soc. Jpn. {\bfseries 78} (2009) 013711.

\bibitem{lankau}
A.~Lankau, K.~Koepernik, S.~Borisenko, V.~Zabolotnyy, B.~B\"uchner, J.~van~den
  Brink, and H.~Eschrig: Phys. Rev. B {\bfseries 82} (2010) 184518.

\bibitem{umezawa}
K.~Umezawa, Y.~Li, H.~Miao, K.~Nakayama, Z.-H. Liu, P.~Richard, T.~Sato, J.~B.
  He, D.-M. Wang, G.~F. Chen, H.~Ding, T.~Takahashi, and S.-C. Wang: Phys. Rev.
  Lett. {\bfseries 108} (2012) 037002.

\bibitem{hajiri}
T.~Hajiri, T.~Ito, R.~Niwa, M.~Matsunami, B.~H. Min, Y.~S. Kwon, and S.~Kimura:
  Phys. Rev. B {\bfseries 85} (2012) 094509.

\bibitem{skornyakov}
S.~Skornyakov, D.~Novoselov, T.~Gürel, and V.~Anisimov: J.~Exp. Theor. Phys.
  {\bfseries 96} (2012) 118.

\bibitem{gslee}
G.~Lee, H.~S. Ji, Y.~Kim, C.~Kim, K.~Haule, G.~Kotliar, B.~Lee, S.~Khim, K.~H.
  Kim, K.~S. Kim, K.-S. Kim, and J.~H. Shim: Phys. Rev. Lett. {\bfseries 109}
  (2012) 177001.

\bibitem{johannes}
M.~D. Johannes and I.~I. Mazin: Phys. Rev. B {\bfseries 77} (2008) 165135.

\bibitem{dong}
J.~Dong, H.~J. Zhang, G.~Xu, Z.~Li, G.~Li, W.~Z. Hu, D.~Wu, G.~F. Chen, X.~Dai,
  J.~L. Luo, Z.~Fang, and N.~L. Wang: Europhys.~Lett. {\bfseries 83} (2008)
  27006.

\bibitem{yin}
Z.~Yin, K.~Haule, and G.~Kotliar: Nature Mater. {\bfseries 10} (2011) 932.

\bibitem{ferber}
J.~Ferber, K.~Foyevtsova, R.~Valent\'i, and H.~O. Jeschke: Phys. Rev. B
  {\bfseries 85} (2012) 094505.

\end{thebibliography}
%\begin{thebibliography}{9}
%\bibitem{jpsj} The abbreviation for JPSJ must be ``J. Phys. Soc. Jpn." \note{in the reference list}.
%\bibitem{instructions} More abbreviations of journal titles are listed in ``Instructions for Preparation of Manuscript".
%\end{thebibliography}

\end{document}